# Implementing Bayesian inference on a stochastic $CO_2$-based grey-box model for assessing indoor air quality in Canadian primary schools

Shujie Yan[1], Jiwei Zou[1,2], Chang Shu[2], Justin Berquist[2], Vincent Brochu[3], Marc Veillette[3], Danlin Hou[1], Caroline Duchaine[3], Liang (Grace) Zhou[2], Zhiqiang (John) Zhai[4], Liangzhu (Leon) Wang[1*]

[1]Dept. of Building, Civil & Environmental Engineering, Concordia University 1455 de Maisonneuve Blvd. West, Montreal, Quebec, Canada

[2]Construction Research Centre, Engineering Division, National Research Council of Canada, M-24, 1200 Montreal Road, Ottawa, Ontario, Canada

[3]Département de biochimie, de microbiologie et de bio-informatique, Faculté des sciences et de génie, Université Laval, Québec, Canada

[4]Department of Civil, Environmental and Architectural Engineering, University of Colorado, Boulder, USA

Corresponding author: Liangzhu (Leon) Wang, Dept. of Building, Civil & Environmental Engineering, Concordia University, Quebec, Canada  leon.wang@concordia.ca

**Abstract**

The COVID-19 pandemic brought global attention to indoor air quality (IAQ), which is intrinsically linked to clean air change rates. Estimating the air change rate in indoor environments, however, remains challenging. It is primarily due to the uncertainties associated with the air change rate estimation, such as pollutant generation rates, dynamics including weather and occupancies, and the limitations of deterministic approaches to accommodate these factors. In this study, Bayesian inference was implemented on a stochastic $CO_2$-based grey-box model to infer modeled parameters and quantify uncertainties. The accuracy and robustness of the ventilation rate and $CO_2$ emission rate estimated by the model were confirmed with $CO_2$ tracer gas experiments conducted in an airtight chamber. Both prior and posterior predictive checks (PPC) were performed to demonstrate the advantage of this approach. In addition, uncertainties in real-life contexts were quantified with an incremental variance $\sigma$ for the Wiener process. This approach was later applied to evaluate the ventilation conditions within two primary school classrooms in Montreal. The Equivalent Clean Airflow Rate (ECAi) was calculated following ASHRAE 241, and an insufficient clean air supply within both classrooms was identified. A supplement of 800 cfm clear air delivery rate (CADR) from air-cleaning devices is recommended for a sufficient ECAi. Finally, steady-state $CO_2$ thresholds ($C_{limit}$, $C_{target}$, and $C_{ideal}$) were carried out to indicate when ECAi requirements could be achieved under various mitigation strategies, such as portable air cleaners and in-room ultraviolet light, with CADR values ranging from 200 to 1000 cfm.

1. Introduction

Since the COVID-19 pandemic started, more than 775.3 million individuals worldwide have been infected, and approximately 7.0 million deaths have been attributed to the disease as of April 7, 2024 [1]. The pandemic has significantly highlighted public concerns about maintaining healthy indoor environments to limit the spread of virus-laden respiratory aerosols [2, 3]. As hygiene and self-protective measures have eased, there are fewer people wearing masks and maintaining social distancing in public spaces. The respiratory diseases, which include not only SARS-CoV-2 but also influenza, respiratory syncytial virus (RSV), etc., would continue to pose health threats [4]. In daily life settings, classrooms in schools are particularly vulnerable [5], where face-to-face interactions are inevitable and frequent. To mitigate health effects from respiratory infections, ensuring sufficient clean air ventilation plays an essential role. Effective ventilation can significantly dilute aerosol concentrations and reduce the quantity of inhaled infectious pathogens. Consequently, assessments of indoor air quality (IAQ) and, more specifically, characterizations of ventilation in schools have become more crucial than ever.

Carbon dioxide ($CO_2$), which serves as an indicator of indoor ventilation conditions, is recommended for managing the risk of airborne transmission [6]. This is because the indoor $CO_2$ level could reflect the outdoor ventilation rate per person, provided that information on occupancy and specific space ventilation requirements is available [7]. During occupants' exposures, indoor $CO_2$ levels will gradually increase until equilibrium is achieved. Poor ventilation conditions will elevate the steady-state $CO_2$ levels, causing them to exceed the recommended $CO_2$ metrics. Meanwhile, its concentration can be conveniently measured with portable low-cost sensors installed in classrooms. St-jean et al. [8] found elevated $CO_2$ levels in 21 day-care centers (DCCs) in Montreal. Andamon et al. [9] reported the elevated $CO_2$ concentration in 10 classrooms of a secondary school in Victoria, Australia. In response to the COVID-19 pandemic, the province of Quebec, Canada, equipped all kindergarten, elementary, high school, vocational, and adult education classrooms with $CO_2$ sensors to monitor indoor air quality and improve ventilation conditions [10]. The widespread installation of $CO_2$ sensors has facilitated enhanced monitoring of indoor ventilation conditions in classrooms.

In addition, there are several $CO_2$-based ways to determine ventilation rates from field measurements: steady-state, decay, build-up, and transient mass-balance approach. Andamon et al. [9] used average peak $CO_2$ concentrations as steady-state values to estimate ventilation rates for investigated classrooms. Kabirikopaei et al. [11] estimated ventilation rates for 220 classrooms in the Midwestern region of the US using three methods (steady-state, decay, and build-up) and found that $CO_2$ readings can contribute the most uncertainty. Batterman [12] suggested that the transient mass balance method can provide the most accurate results when occupancy is available.

While these traditional approaches have been widely used for indoor ventilation rate evaluations, there are several limitations. Firstly, most of these approaches adopted deterministic $CO_2$ mass-balance equations, assuming parameters in the model to be constants. In practical settings, however, multiple sources of uncertainties may exist in the room. Secondly, in real-time $CO_2$ measurements, occupancy data are often unavailable. Inaccurate estimates of occupancy can introduce biases into final evaluation results. Meanwhile, current

$CO_2$ metrics are established only for ventilation standards such as ASHRAE 62.1 [13], and the metrics for managing the long-range transmission of airborne aerosols are yet to be determined.

To capture the uncertainties in the $CO_2$-based ventilation evaluation process, a grey-box model [14] can be used, which usually integrates a partial theoretical structure with data to complete the model. Compared with white-box (e.g., physically-based) and black-box (e.g., data-driven) models, the grey-box model can be structured with physical knowledge, and the parameters are estimated with the measured data from the system. The stochastic grey-box model often includes stochastic items to account for uncertainties and variability in the system [15]. The randomness of input parameters will allow for the consideration of uncertain components [16] such as measurement errors, fluctuations in the system, unmodelled parameters, etc.

Haghihat et al. [17] introduced a predictive stochastic model for indoor air quality in 1988, allowing the incorporation of inputs as random variables within the stochastic differential equation (SDE) model. The model can capture variability in predictions of contaminant concentrations. The moment equations for mean, variance, and skewness were given based on stochastic Itô calculus [17]. It was indicated that the 'white noise' term not only described the system randomness but also provided a unique and satisfactory solution. It is worth noting that the solution of the SDE model is an Itô stochastic process with the Markov property and the strong Markov property, which enables future predictions to rely only on the current state [18]. Marcel et al. [16] proposed a predictive control approach to model the $CO_2$ concentrations using a grey-box model, in which SDE equations were established based on tracer-gas mass balance. The study suggests that the parametrization of the model was suitable and applicable, and the prediction tends to be more accurate than traditional deterministic approaches. Until now, studies that attempted to interpret indoor ventilation conditions using the grey-box model are still rare [15, 16].

Parameter estimation plays a key role in developing a stochastic grey-box model. Contemporary improvements in computational power have substantially enhanced Bayesian inference, making it a robust tool for precise parameter estimations, uncertainty quantification, and effective incorporation of prior knowledge. Many previous efforts have been delivered to apply Bayesian inference to interpret parameters in IAQ models. Wang et al. [19, 20] applied the Bayesian approach to a source-detector relationship established from CFD simulations of flow fields in indoor spaces and underground utility tunnels for estimating source parameters (leakage rate and location). Septier et al. [21] proposed a Bayesian inference procedure on inverse dispersion modeling to solve the challenging source term estimation (STE) problem. The Gaussian assumption was made for the source emission rate for its satisfactory performance in practice, even though the emission rate cannot take negative values. To assess ventilation conditions with $CO_2$ meters in primary schools, Hou et al. [22] applied a Bayesian inference approach to indoor $CO_2$ concentration models. This study identified the outdoor ventilation rate, $CO_2$ generation rate, and occupancy level as the most sensitive variables to indoor $CO_2$ levels. Rahman et al. [23] developed an approach to estimate the occupancy distribution in a mechanically ventilated multi-room office using Bayesian inference. The $CO_2$ concentration, simulated by the CONTAM program, was taken as input for the investigation under the circumstances with and without a 5% random noise considered for uncertainty. The study found a significant increase in the RMSE in estimating occupancy as the sensor measurement uncertainty increases. By applying the moving-average filtering method, the RMSE on estimation was reduced, however it became insensitive to the abrupt occupancy change. It was suggested that the Bayesian inference would be more powerful in solving inverse problems if it could handle realistic data including noises. However, existing literature reveals a noticeable scarcity of interpreting parameters from stochastic models with Bayesian inference.

To summarize, most traditional ventilation evaluation approaches utilize deterministic approaches that cannot accommodate real-life uncertainties. The accuracy of these approaches would rely on how the real situations approach the idealized assumptions, the accuracy and comprehensive collection of inputs, the correct and comprehensive model development, and no disturbances during the measurements, etc. Idealized situations seldom happen in reality, so quantifying uncertainties is essential.

In this study, we employed Bayesian inference on a $CO_2$-based grey-box SDE model for assessing ventilation conditions for two classrooms in Montreal with $CO_2$ field measurement data. The methodology and data used in this study will be introduced in Section 2. In Section 3, the modeling results will be presented and discussed. In Section 3.1, a prior sensitivity analysis was conducted on the model. In Section 3.2, model validations from an airtight chamber are demonstrated for ventilation rate (Section 3.2.1) and $CO_2$ emission rates (Section 3.2.2). The posterior predictive checks (PPC) and noise-level estimation results are discussed in Sections 3.2.3 and 3.2.4, respectively. Section 3.3 illustrates the case study outcomes, evaluating the ventilation conditions and providing ECAi across three seasons: Spring (March to May), Autumn (September to November), and Winter (December to February). The $CO_2$ threshold necessary to meet ECAi requirements from ASHRAE 241 was estimated for the classrooms to manage the long-range aerosol exposures. The conclusions of this study are presented in Section 4.

## 2. Methodology

The methodology used in this study implemented Bayesian inference on a stochastic $CO_2$-based grey-box model to interpret parameters and quantify the modeling uncertainties. In Section 2.1, the stochastic $CO_2$-based grey-box SDE model will be introduced, followed by the principles of Bayesian inference to be explained in Section 2.2. The validation and PPC process are detailed in Section 2.3. The model development process is illustrated in Fig. 1.

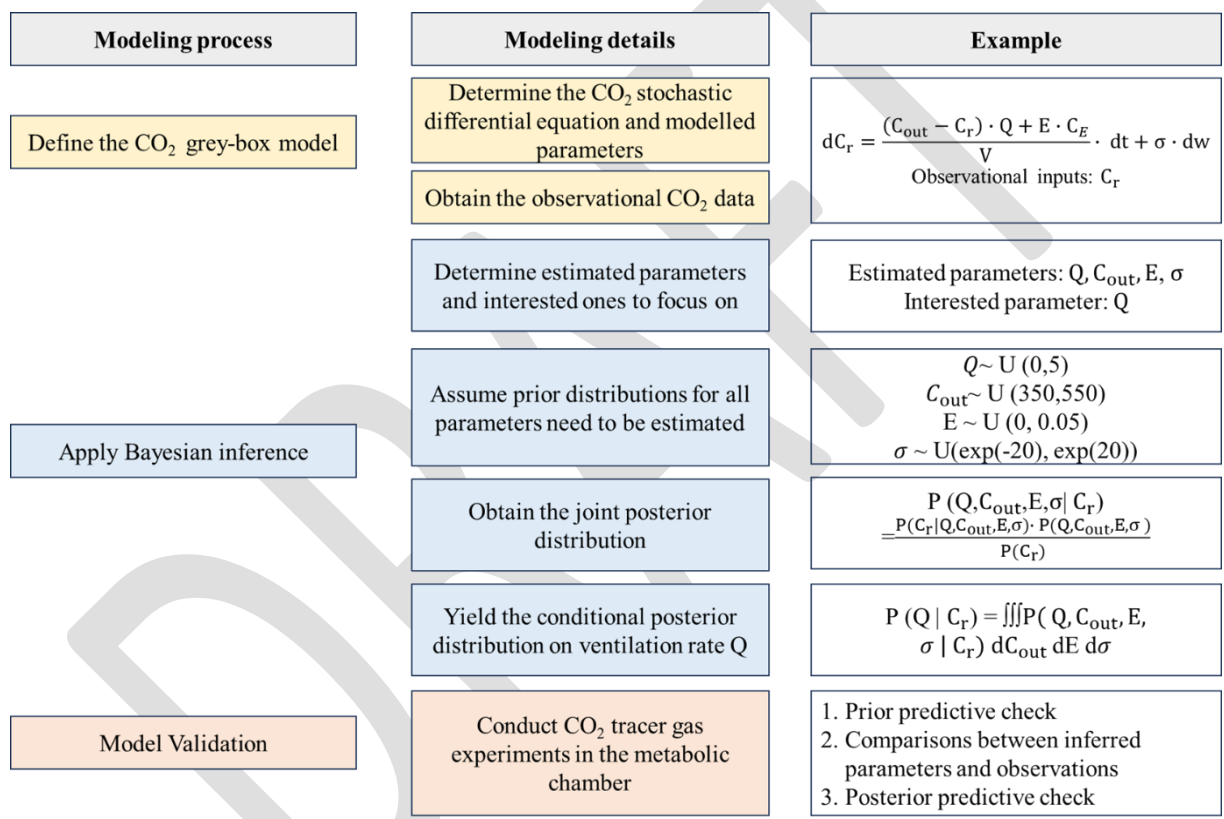

Fig. 1 The model development process

### 2.1 Stochastic $CO_2$-based grey-box model

When establishing models for IAQ problems, the deterministic mass-balance equation is often established for the contaminants as an ODE. If the randomness of some coefficients is allowed, it will become an SDE, and it tends to be more realistic to the real-life problems of interest. The establishment of a stochastic $CO_2$ grey-box model is composed of two components [18]: a drift component, which represents the deterministic description of the system, and a diffusion term, which represents the stochastic or random evolution of the system. The diffusion component captures the variability or uncertainty in the system's behavior attributable to random forces or noise.

The drift components derived from the traditional deterministic $CO_2$ mass-balance model, or ODE, are represented by Eq. 1:

$$V \frac{dC_r}{dt} = (C_{out} - C_r) \cdot Q + E \cdot C_E \quad \quad 1)$$

Where V is the room volume (L); $C_r$ is the $CO_2$ concentration in the room (ppm); $C_{out}$ is the $CO_2$ concentration of outdoor air or the ventilation flows (ppm); Q is the ventilation rate (L/s); E is the total volumetric $CO_2$ generation rate in the room (L/s); $C_E$ is the conversion factor from volumetric concentration to ppm, which equals to $10^6$.

The diffusion component can be considered as a 'white noise' term added to Eq. 1 accounting for the uncertainty associated with unknown model inputs and other noises in the system. The stochastic $CO_2$ grey-box model could, therefore, be expressed as an SDE as shown in Eq. 2:

$$dC_r = \frac{(C_{out} - C_r) \cdot Q + E \cdot C_E}{V} \cdot dt + \sigma \cdot dW_t \quad \quad 2)$$

Where $W_t$ is a Wiener process, which is also known as Brownian motion, is a continuous-time stochastic process that has been widely explored in physics, economics, and applied mathematics [24], and $\sigma$ is the incremental variance in the Wiener process (ppm/$\sqrt{dt}$).

In this study, the Euler-Maruyama method was used for discretization [25], which provides an approximate solution to the SDE equation over discrete time steps. The Euler-Maruyama approximation is provided through Eq. 3:

$$\Delta C_r = (C_{out} - C_r) \cdot Q + E \cdot C_E)/V \cdot \Delta t + \sigma \cdot \sqrt{\Delta t} \cdot Z \quad \quad 3)$$

Where $\Delta C_r$ is the change in $C_r$ over the time step $\Delta t$; Z is a standard normal random variable (from a normal distribution with mean 0 and variance 1).

In this study, the inclusion of the 'white noise' component assists in quantifying the uncertainty levels in model predictions. Such uncertainties might arise from various sources, including air turbulence, systematic measurement errors, inaccuracies in estimating modeled parameters, effects of unmodeled parameters, variability in the distribution of occupants within a room, and the positioning of sensors. The component also covers factors that are not accounted for in traditional deterministic models, which could lead to discrepancies between the models and the observations. Assumptions made in this study to use this model are: The room is a well-mixed single-zone space; The estimated parameters are assumed to be constant throughout the evaluation duration; The differences of density between indoor and outdoor air are ignored.

**2.2 Bayesian inference**

Bayesian inference is a powerful tool for quantifying uncertainty in estimated model parameters [26-29]. It considers the inferred parameters as random variables with prior information, and then a likelihood function (based on the measurement data) is used to update prior distribution following Bayes theorem [30]. The updated results are the posterior distributions, which are the new beliefs of the interested variables. In recent years, with the advancement in computational capabilities and the development of Markov Chain Monte Carlo (MCMC) algorithms such as Metropolis-Hastings, Gibbs sampling, and Hamiltonian Monte Carlo, an increasing number of studies in the built environment field started to utilize this approach for parameter inferences in established models [29].

In the defined stochastic $CO_2$ grey-box model, there could be multiple variables required to be estimated. In this study, parameters to be estimated in the model are ventilation rate Q, outdoor $CO_2$ concentration $C_{out}$, generation rate E, and incremental variance $\sigma$ in the Wiener process. By placing prior distributions on all estimated parameters and updating these beliefs, a joint posterior distribution for the entire set of parameters can be obtained. If one parameter is selected as the interested parameter, its marginal posterior distribution will need to be carried out, and the remaining parameters will be regarded as nuisance parameters. With the MCMC algorithms, samples could be drawn from the joint posterior distribution to estimate the conditional posterior distribution of interested parameters. For example, the estimation process for the ventilation rate Q is illustrated as follows:

Step 1: Assume prior distributions for estimated parameters;

Step 2: Obtain the joint posterior distribution of all estimated parameters;

Step 3: Yield the conditional posterior distribution on ventilation rate Q

The prior assumptions are the prior beliefs of the estimated parameters. An example of the prior settings is illustrated in Fig. 2. The impact of informative priors and non-informative priors (flat priors) on posterior distributions will be evaluated in this study. For example, for the outdoor $CO_2$ concentrations, a uniform distribution is assumed to be in the range of 350 ppm to 550 ppm [31]. If there is some information about the outdoor $CO_2$ level available, a normal distribution could then be assumed with a specific mean value and variance level.

Then, the probability of the estimated parameters could be inferred based on the prior distributions estimated for them. The likelihood of the estimated parameters given the measured data Cr ($CO_2$ indoor concentration) is demonstrated as follows in Bayes's theorem (Eq. 4):

$$P(Q, C_{out}, E, \sigma | Cr) = \frac{P(Cr | Q, C_{out}, E, \sigma) \cdot P(Q, C_{out}, E, \sigma)}{P(Cr)} \qquad 4)$$

Where $P(Cr | Q, C_{out}, E, \sigma)$ is the likelihood probability that measurement data Cr occurs given the prior information, $P(Q, C_{out}, E, \sigma)$ is the joint prior distribution of parameters $Q, C_{out}, E$, and $\sigma$, and $P(Cr)$ is the probability of seeing the measurement results, which is a normalized constant.

After obtaining the joint posterior distribution $P(Q, C_{out}, E, \sigma | Cr)$, samples could be drawn from this joint posterior distribution for the nuisance parameters for $C_{out}, E, \sigma$ to estimate the conditional posterior distribution on ventilation rate $Q$, which is $P(Q | C_{out}, E, \sigma, Cr)$. The process for estimating other parameters follows the same procedure.

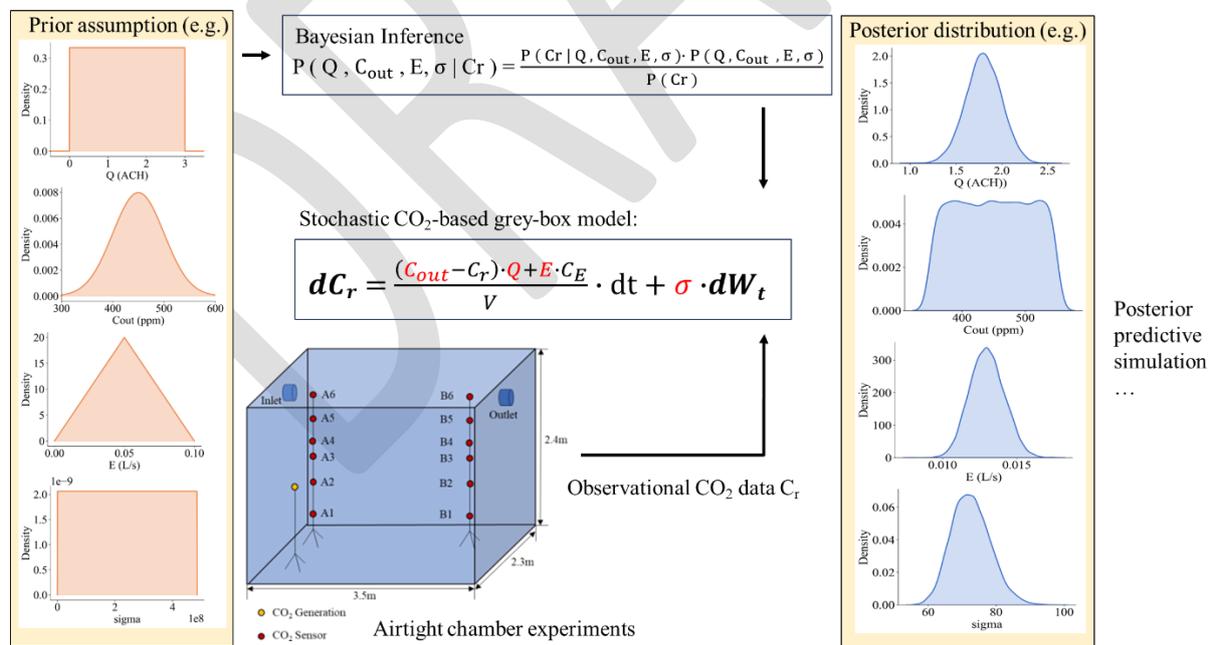

Fig. 2 The illustration of modeling inputs and outputs

In this study, the likelihood function will be estimated using the Euler-Maruyama approximation [25]. The Markov chain Monte Carlo (MCMC) method was applied for reconstructing the uncertain parameters. Given the prior distribution and likelihood, the posterior distribution could be obtained. Five thousand draws from the No-U-Turn Sampler (NUTS) algorithm were performed on two chains to sample the parameter intervals using MCMC. The 'burn-in' was set at 500 to help the Markov chain start near the center of equilibrium distribution. The Bayesian stochastic modeling process was established in the Python module PyMC [32].

## 2.3 Validations and PPC evaluations

Experimental validations and PPC evaluations were carried out to assess the performance and validity of the Bayesian inference results on the stochastic $CO_2$-based grey-box model. Experimental validations are designed to help confirm the model estimation accuracy on ventilation rates and $CO_2$ emission rates. Meanwhile, the PPC will help evaluate how well the developed model fits the observed data. It is conducted to assess the goodness of fit and adequacy. If the fitness is good, it indicates that the model can generate data in patterns similar to those observed.

### 2.3.1 Experimental validations

$CO_2$ tracer gas experiments were conducted in an airtight chamber located at the University Institute of Cardiology and Pneumology of Quebec - Université Laval (IUCPQ - ULaval). The dimension of the airtight chamber is 2.3 m (width) × 3.5 m (length) × 2.4 m (height). The inlet and outlet of the mechanical system are at the top of the chamber with a diameter of 5.1 cm (2 in). Two sensor trees were set up in the chamber, and each of the trees is equipped with mounts at six different heights [33]: 0.6 m, 1.1 m, 1.5 m, 1.7 m, 2 m, and 2.3 m. The sensors were established to confirm the uniform distributions of $CO_2$ inside the chamber. Each of the mounts carries six sets of sensors measuring $CO_2$ (Vaisala - GMP252), air temperature and relative humidity (Vaisala - HMP110), and air velocity (SWEMA 03+), respectively. Details of the sensor specifications are listed in Table 1. The $CO_2$ was generated through the $CO_2$ tank outside the chamber (Fig. 3 (b)), and a mass flow controller was used to control the generation rate.

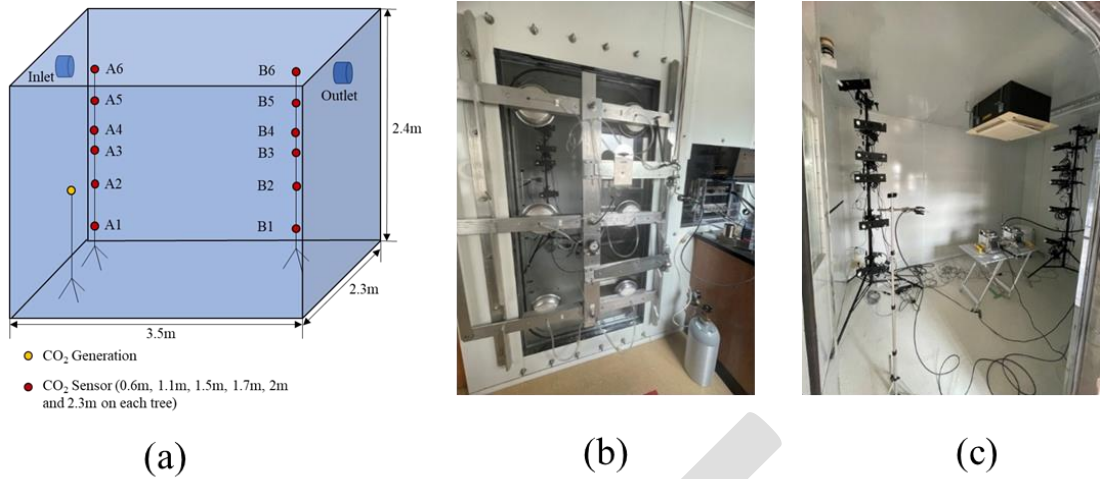

Fig. 3 Experimental set-ups of the airtight chamber; (a) chamber dimensions and designated measurement locations (red point- $CO_2$ sensors, yellow point-$CO_2$ generation location); (b) detailed view of the gas injection and sealing mechanisms; (c) experimental set-ups in the chamber

Table 1 Sensor specification details

| Reading Type | Sensor Name | Measurement range | Accuracy | Sampling Frequency |
|---|---|---|---|---|
| $CO_2$ | Vaisala GMP252 | 0 – 10000 ppm | ± 40 ppm | 0.05 Hz |
| Relative Humidity and Temperature | Vaisala HMP110 | 0–100 % RH<br>-40 – 80 °C | ± 1.5 % RH<br>± 0.2 °C | 0.05 Hz |
| Airspeed | Swema 03+ | 0.05 – 3 m/s | ± 0.03 m/s | 100 Hz |

The experiments were completed in two sessions: concentration decay and constant injection. In decay measurements, three ventilation conditions (Test 1: Ventilation mode 1- 1.9 ± 0.03 ACH; Test 2: Ventilation mode 2 - 1.51 ± 0.02 ACH; Test 3: Ventilation mode 3 - 0.53 ± 0.01 ACH) were measured. Due to the limited conditions for directly measuring the supply airflows, the referenced ventilation rates for the three different ventilation conditions were calculated from the $CO_2$ decay approach. A fan was operated during the initial mixture period. The $CO_2$ injection stops upon the peak and stabilization of $CO_2$ concentration, and the concentration is recorded throughout the subsequent decay period. During the constant $CO_2$ injection tests, two distinct $CO_2$ generation rates, 0.8 L/min, and 1.6 L/min were examined with and without fan operation (Test 4: 0.8 L/min, fan-off; Test 5: 0.8 L/min, fan-on; Test 6: 1.6 L/min, fan-off; Test 7: 1.6 L/min, fan-on). These measurements were conducted under the chamber's Ventilation mode 1. After the $CO_2$ tracer gas experiments, the measured $CO_2$ data are used as observational data for the model, as illustrated in Fig. 2.

### 2.3.2 PPC evaluation

PPC is a useful way of assessing the model and determining if it fits the data directly. Specifically, to check the model's fit, the simulated values were drawn from posterior predictive distributions, and the samples were compared with the observed data. If the proposed model fits, the regenerated simulations from the model should resemble the observations and no major discrepancy would be observed. Traditionally, the previous classical approaches mainly focus on various goodness-of-fit tests, comparing a tested statistic derived from observed data to its distribution under the null hypothesis. Unlike the traditional p-value in frequentist statistics, the Bayesian p-value helps to evaluate how well a Bayesian model describes the observed data. The Bayesian p-value is defined in Eq. 5.

$$\text{Bayesian p-value} \triangleq p\,(T_{sim} \geq T_{obs} \mid Cr) \quad \quad 5)$$

Where $T_{sim}$ is the simulated statistic, $T_{obs}$ is the statistic for observations, and $Cr$ is the conditions of the observations. A Bayesian p-value close to 0.5 would suggest a good fit, indicating that the observed data appears typical of the data predicted by the model. When values close to 0 or 1, however, would indicate a poor fit, suggesting that observations are impossible under the model. In this study, the target test statistic is the posterior mean value.

## 3 Results and discussion

### 3.1 Prior sensitivity analysis on inferred parameters

A crucial component in the Bayesian modeling and inference process is the prior distribution, which represents our initial assumptions or knowledge about unknown model parameters. In Bayesian analysis, this prior distribution is subsequently combined with the likelihood, which is the probability that observation occurs, given the parameters, to obtain the posterior distribution. The posterior distribution thus reflects an updated belief of the parameters, incorporating both our prior knowledge and the new evidence. Therefore, the prior assumptions may have a significant impact on the posterior estimates of mean, bias, quantiles, etc. In this section, a prior sensitivity analysis was conducted for the stochastic $CO_2$ grey-box model to help assess whether the inferred results are influenced by the prior assumption settings.

The prior sensitivity analysis was conducted for four hyperparameters in the model: ventilation rate Q, outdoor $CO_2$ concentration level $C_{out}$, $CO_2$ emission rate E, and incremental variance $\sigma$. In addition, two types of priors are chosen for the investigation: vague proper prior and informative prior. There are two main objectives for this prior sensitivity analysis. The first is to investigate whether different prior assumptions will influence the parameter of interest, for example, ventilation rate Q and $CO_2$ emission rate E. The second is to investigate whether the prior assumption of nuisance parameters, which are the parameters that are not of direct interest, would play an important role in estimating the interested parameters. The nuisance parameter investigated in this study is the outdoor $CO_2$ concentration level $C_{out}$.

The prior sensitivity analysis for one case of this study will be demonstrated here, using the observational data from one constant injection experimental test conducted in the airtight chamber (Test 4). The observational data for this investigated scenario is illustrated in Fig. 4. The metabolic chamber was set at its Ventilation mode 1 (ACH = 1.9 ± 0.03), and $CO_2$ tracer gas was constantly injected into the chamber at a rate of 0.8 L /min (0.013 L/s). The information on the investigated priors is listed in Table 2.

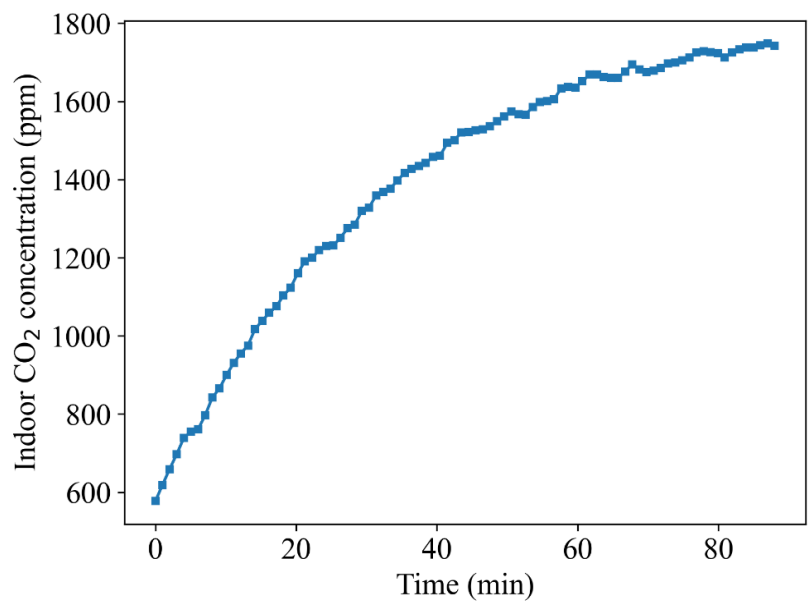

Fig. 4 The investigated test for prior sensitivity analysis
(Test 4 in the airtight chamber)

Table 2 Evaluated priors for inferred parameters

| Parameter | Unit | Default Prior | Vague Proper Prior | Informative Prior |
|---|---|---|---|---|
| Q | ACH | U (0,3) | U (0,10) | N (2, 0.2) |
| $C_{out}$ | ppm | U (350,550) | U (350,550) | U (396,416) N (400,20) |
| E | L/s | U (0,0.05) | U (0,0.1) | N (0.013,0.005) |

Note. U = Uniform; N = Normal

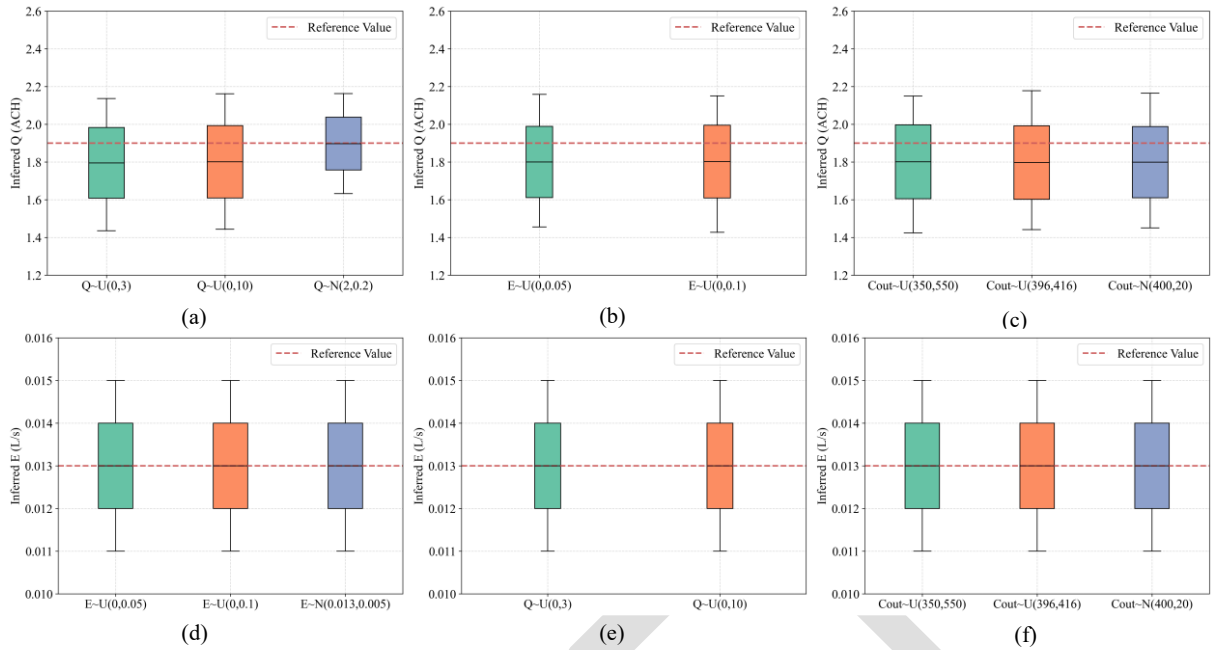

Fig. 5 Prior sensitivity analysis on estimated parameters; U = Uniform; N = Normal; When investigating a specific prior assumption for a given parameter, default priors will be applied to other parameters; (a) – (c): The influence of prior assumptions of Q, E, and $C_{out}$ on Q; (d) – (f): The influence of prior assumptions of Q, E, and $C_{out}$ on E

As depicted in Fig. 5, the influence of prior assumptions on parameter inference was investigated within the stochastic $CO_2$ grey-box model. Specifically, we focus on the ventilation rate (Q) and the $CO_2$ emission rate (E), aiming at elucidating the model's sensitivity and robustness to varying prior assumptions. The robustness of the model estimations for ventilation rate Q, under varying prior assumptions, is illustrated in Fig. 5 (a) - (c). It is observed that the adoption of a vague uniform prior (Q ~ U (0,10)) does not introduce significant deviations from the results generated by the default uniform prior (Q ~ U (0,3)). This indicates that a broader assumption range will not influence final evaluations. As a result, both of the two prior assumptions can be considered proper priors in this study. Furthermore, the adoption of an informative prior (Q ~ N (2, 0.2)) could enhance the model's prediction performance on Q (Fig. 5 (a)), as would be expected. Besides, the changes in prior assumptions regarding the $CO_2$ emission rate (E) and outdoor $CO_2$ concentration level ($C_{out}$) demonstrate negligible impacts on the model's inference performance for Q, as illustrated in Fig. 5 (b) and Fig. 5 (c). It suggests the model's insensitivity to these prior assumptions, thereby reinforcing the robustness of its estimations on the parameter Q.

Similar findings were also observed for the estimations of $CO_2$ emission rate E, as shown in Fig. 5 (d). The model achieves good accuracy in estimating E, even under the vague prior (E ~

U (0, 0.1)). This capacity is further demonstrated in Fig. 5 (e) and Fig. 5 (f), where the estimation accuracy for E was not affected by the varying priors for Q and $C_{out}$.

In conclusion, employing Bayesian inference on the stochastic $CO_2$ grey-box model exhibits good robustness to the variations in prior assumptions for both Q and E. The model could maintain its predictive accuracy across a range of prior assumptions, from vague proper to highly informative prior. This evaluation results prove the rationality of the prior assumptions made for the model.

### 3.2 Model validation

In this section, the accuracy of the inferred parameters (Q and E) was further validated, adopting the verified default prior assumptions listed in Table 2. Two types of $CO_2$ tracer gas tests were used as observational data inputs in this model validation process, distinguished by the $CO_2$ generation situations: with or without $CO_2$ release. Detailed information for the validation scenarios is listed in Table 3 and Table 4. The validation results for the model's estimation performance are presented in the following subsection.

### 3.2.1 Validation of the inferred ventilation rate Q

One of the main purposes of this study is to make reasonable inferences of ventilation rate Q from the indoor $CO_2$ measurement with the proposed approach. The model estimation performance for Q was compared with the ventilation rate obtained from standard tracer-gas decay tests. For the concentration decay tests, there were three ventilation modes investigated. The comparison results are illustrated in Table 3 and Fig. 6. For the concentration decay test, the difference is less than 5% for the three ventilation modes.

When the $CO_2$ release was considered in the model (constant injection test), the relative errors for the two release conditions 0.8 L/min (0.013 L/s) and 1.6 L/min (0.026 L/s) were 5.7% and 3.1% under fan-off situations. For the "fan on" scenarios, the differences increase to 12.6% and 11% for the two release conditions. The reasons for this increase in differences are probably due to the fan actively circulating air in the small chamber, leading to a higher actual ventilation rate. It should be noted that there was no obvious difference in the $CO_2$ measurements observed at the twelve sensors in the two sensor trees, thus the influence of non-uniformity could be ignored. To summarize, the proposed approach in this study can make reasonable ventilation rate estimations for both decay and constant injection scenarios.

Table 3 Validations for inferred ventilation rate Q

| $CO_2$ Tracer gas measurements | Test number | Experimental conditions | Estimated Q (ACH) | | Experimental Q (ACH) | | Relative Error (%) |
|---|---|---|---|---|---|---|---|
| | | | mean | sd | mean | sd | |
| Concentration decay | Test1 | Ventilation 1 | 1.90 | 0.03 | 1.91 | 0.03 | 0.5% |
| | Test2 | Ventilation 2 | 1.52 | 0.02 | 1.51 | 0.04 | 0.6% |
| | Test3 | Ventilation 3 | 0.51 | 0.01 | 0.53 | 0.03 | 3.8% |
| Constant injection | Test4 | Ventilation1, $CO_2$ release =0.013 L/s (0.8 L/min), fan off | 1.80 | 0.20 | 1.91 | 0.03 | <u>5.7%</u> |
| | Test5 | Ventilation1, $CO_2$ release =0.013 L/s (0.8 L/min), fan on | 2.15 | 0.26 | 1.91 | 0.03 | *12.6%* |
| | Test6 | Ventilation1, $CO_2$ release =0.026 L/s (1.6 L/min), fan off | 1.85 | 0.31 | 1.91 | 0.03 | <u>3.1%</u> |
| | Test7 | Ventilation1, $CO_2$ release =0.026 L/s (1.6 L/min), fan on | 2.12 | 0.06 | 1.91 | 0.03 | *11.0%* |

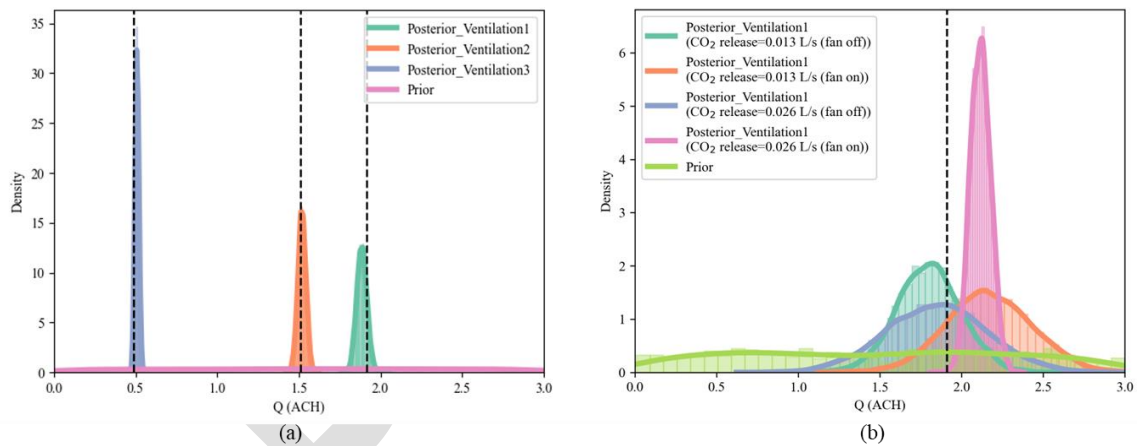

Fig. 6 Posterior distribution on inferred ventilation rate Q; a) Inferred ventilation rate under three ventilation conditions from $CO_2$ decay tests; b) Inferred ventilation rate under normal ventilation conditions from $CO_2$ constant injection tests; The black dashed line indicates the referenced value measured from the airtight chamber

### 3.2.2 Validation of the inferred $CO_2$ emission rate E

In addition to the ventilation rate Q, the $CO_2$ emission rate E was another parameter of interest, which can be inferred from the model simultaneously. The interpretation of the $CO_2$ emission rate in the room can help estimate the occupancy based on the $CO_2$ emission rate per person under certain ages and physical activity levels. The comparisons between the inferred $CO_2$ emission rates E and the $CO_2$ emission readings from the mass flow controller are illustrated in Table 4 and Fig. 7.

From Fig. 7, it could be found that the inferred $CO_2$ emission rate E is in good agreement with the measurements obtained from the emission mass flow controller for all tested scenarios. As shown in Table 4, in the "fan off" scenario, the estimation errors for the release rates of 0.8 L/min and 1.6 L/min were 2.3 % and 3.8%, respectively. However, when the fan was turned on, the discrepancies widened, increasing to 12.8 % for 0.8 L/min and 11.5 % for 1.6 L/min. Though the differences increased in the "fan on" conditions when compared with the "fan off" ones, the estimated errors remained in an acceptable range of 5% - 15%.

Table 4 Validations for inferred $CO_2$ emission rates E

| Constant injection | Experimental conditions | $CO_2$ emission rate E (L/s) | | $CO_2$ emission readings from mass flow controller (L/s) | | Relative Error (%) |
|---|---|---|---|---|---|---|
| | | mean | sd | mean | sd | |
| Test 4 | Ventilation 1, $CO_2$ release =0.013 L/s (0.8L/min), fan off | 0.013 | 0.001 | 0.0133 | 0.0001 | 2.3 % |
| Test 5 | Ventilation 1, $CO_2$ release =0.013 L/s (0.8L/min), fan on | 0.015 | 0.001 | 0.0133 | 0.0001 | 12.8 % |
| Test 6 | Ventilation 1, $CO_2$ release =0.026 L/s (1.6L/min), fan off | 0.025 | 0.003 | 0.0267 | 0.0002 | 6.4 % |
| Test 7 | Ventilation 1, $CO_2$ release =0.026 L/s (1.6L/min), fan on | 0.029 | 0.001 | 0.0267 | 0.0002 | 8.6 % |

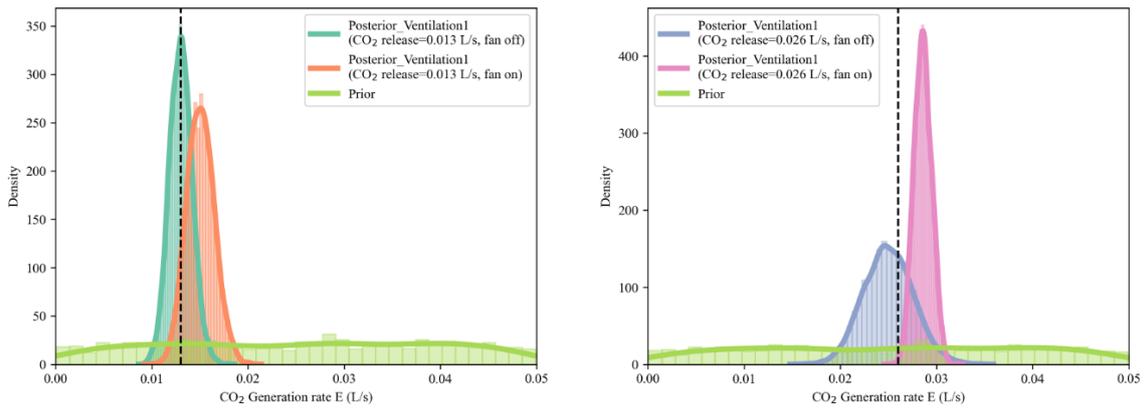

Fig. 7 Posterior distribution on $CO_2$ emission rate E; a) Inferred $CO_2$ generation rate under $CO_2$ release = 0.013 L/s (0.8 L/min); b) Inferred $CO_2$ generation rate under $CO_2$ release = 0.026 L/s (1.6 L/min); The black dashed line indicates the referenced value measured from the airtight chamber

### 3.2.3 PPC evaluation results

The PPC evaluation results for decay and constant injection scenarios are shown in Fig. 8 and Fig. 9 as follows. It suggests that the generated data closely align with the observed data, which further validates the accuracy of the inferred parameters. The Bayesian p-value for the decay and constant injection scenarios fall in the range of 0.37 - 0.40 and 0.63 - 0.66, respectively, all close to 0.50, which suggests a reasonable fit for the model.

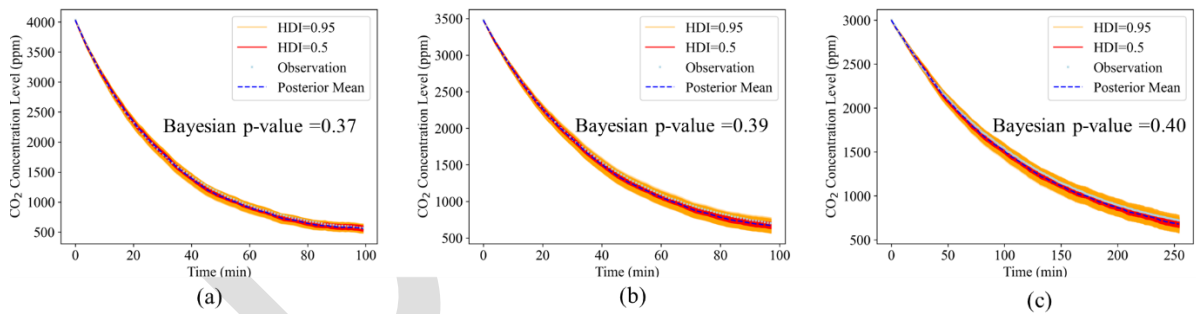

Fig. 8 Posterior predictive simulations for decay scenarios (HDI =Highest Density Interval, indicating the probability that true value drops in this interval) ((a) Ventilation 1, (b) Ventilation 2, (c) Ventilation 3)

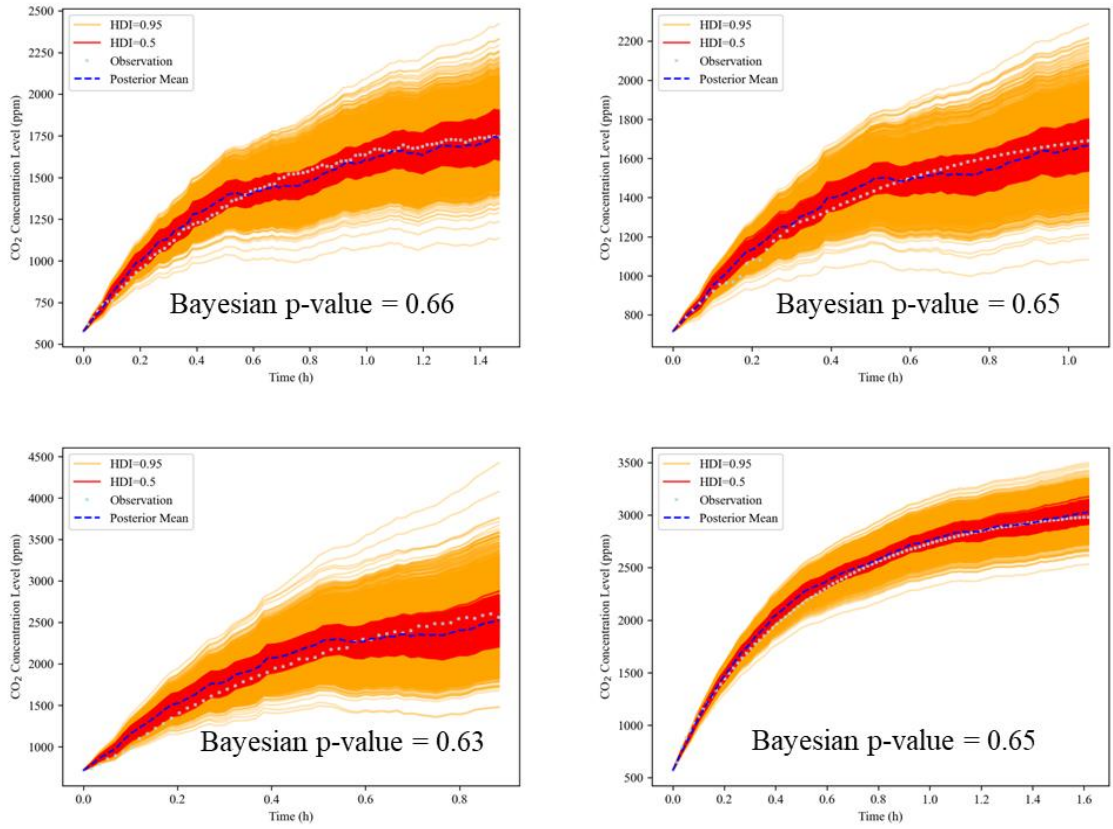

Fig. 9 Posterior predictive simulations for constant injection scenarios (HDI =Highest Density Interval, indicating the probability that true value drops in this interval; scenario details see Table 4)

### 3.2.4 Noise level estimations in $CO_2$ trend predictions

In this section, the posterior means from the constant injection scenarios (Table 5) are taken as inputs for the ODE and SDE $CO_2$ mass-balance models (Eq. 1 and Eq. 2), respectively. One hundred simulations were conducted for the SDE model. The $CO_2$ trend predictions and noise level estimations are illustrated in Fig. 10. The SDE model could capture the variability in the observational data and make reasonable estimations for the $CO_2$ trend. Compared with the traditional ODE model, SDE predictions could consider real-life noise estimations and make unbiased predictions. The posterior distributions estimated for $\sigma$, which is the incremental variance in the Wiener process to scale the magnitude of the random fluctuation, were shown in Table 5. Correspondingly, the noise level estimations are shown in Fig. 10 (b) (d) (f) (h).

For the scenarios with a $CO_2$ release rate of 0.013 L/s, the disturbances in the system illustrated similar magnitudes with $\sigma$ estimations at 72.7 ± 5.8 (fan off) and 75.4 ± 7.1 (fan on), respectively. No significant differences were observed, and the predicted noise levels both fall in the range of -100 ppm to 100 ppm. In scenarios where the $CO_2$ release rate was 0.026 L/s, the disturbances were significantly reduced when the fan was on. For the fan-off condition, the $\sigma$ was estimated to be 157.3 ± 16.9, whereas this estimation dropped to 48.6 ± 3.6 when the fan was turned on. The fan's operation appeared to reduce the variability in the observed data and this effect was not obvious when the $CO_2$ release rate was low. It should be noted that throughout the experiments, the fan was controlled remotely, ensuring its operation was the only altered condition. All other experimental conditions were kept constant during these tests. According to the manufacturers, the sensor measurement errors are ± 40 ppm, which is captured by the scenarios listed in Table 5.

Table 5 Posterior distributions estimated for the incremental variance

| Test number (for constant injection) | Experimental conditions | $\sigma$ (ppm/$\sqrt{h}$) mean | $\sigma$ (ppm/$\sqrt{h}$) sd | Relative ratio to steady-state $CO_2$ level in ppm (mean estimations from one hundred SDE simulations) |
|---|---|---|---|---|
| Test 4 | Ventilation 1, $CO_2$ release =0.013 L/s (0.8 L/min), fan off | 72.7 | 5.8 | ± 1 % |
| Test 5 | Ventilation 1, $CO_2$ release =0.013 L/s (0.8 L/min), fan on | 75.4 | 7.1 | ± 2.2 % |
| Test 6 | Ventilation 1, $CO_2$ release =0.026 L/s (1.6 L/min), fan off | 157.3 | 16.9 | ± 1.1 % |
| Test 7 | Ventilation 1, $CO_2$ release =0.026 L/s (1.6 L/min), fan on | 48.6 | 3.6 | ± 0.9 % |

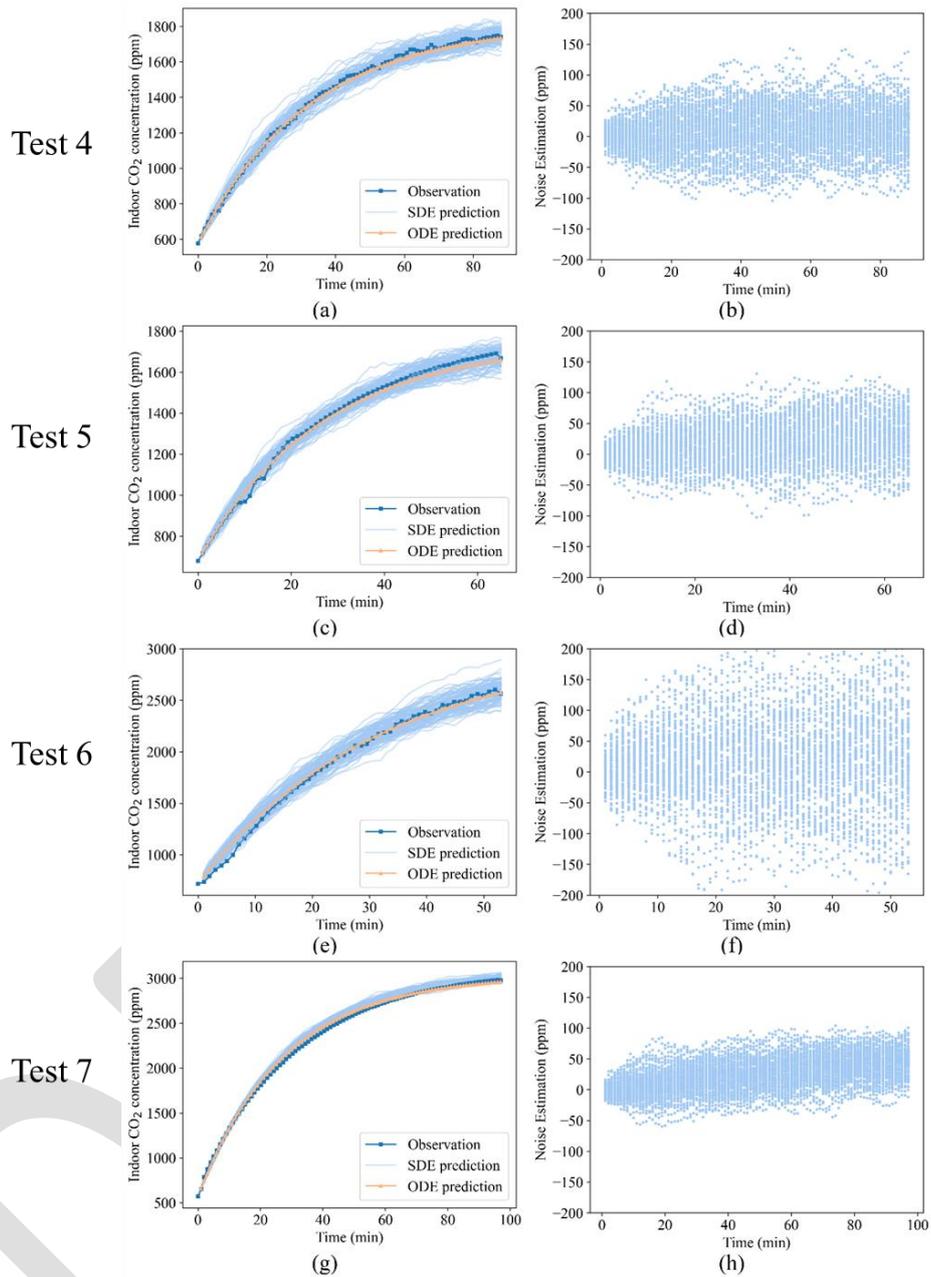

Test 4

Test 5

Test 6

Test 7

Fig. 10 ODE- and SDE-based $CO_2$ trend predictions and noise level estimations

### 3.3 Case Study

Indoor whole-year field measurements of $CO_2$ levels from two Montreal primary schools (from 2020 to 2021) were used to employ the proposed approach in real-life settings. The selected classrooms have a floor area of 9.4 m × 6.6 m (Classroom 1) and 8.8 m × 7.1 m (Classroom 2), respectively, and both are naturally ventilated. The HOBO Bluetooth Low Energy Carbon Dioxide- Temp / RH Data Logger was installed in classrooms at 1.7 meters height on the west internal wall right above the thermostat (1.5 m height). The detailed information on the data logger is listed in Table 6.

Table 6 Detailed information for the HOBO Data Logger

| Reading Type | Measurement range | Accuracy | Resolution |
|---|---|---|---|
| $CO_2$ | 0 – 5000 ppm | ± 50 ppm | – |
| Relative Humidity | 1% – 90% RH | ± 2 % RH | 0.01 % |
| Temperature | -20 – 70 ºC | ± 0.21 ºC | 0.024 °C at 25°C |

Table 7 Measurements information in the classroom

| Classroom | Location | Age | Dimensions (m) | Ventilation Type | Measurement Periods |
|---|---|---|---|---|---|
| Classroom 1 | Montreal | 5-8 | 9.4 × 6.6 ×3.47 | Natural ventilation | 2020/06/22 - 2021/06/21 |
| Classroom 2 | Montreal | 5-8 | 8.8 × 7.1 × 3.2 | Natural ventilation | 2020/08/26 - 2021/08/25 |

Table 7 illustrates the measurement information for the two primary classrooms. One week of weekday data (from Monday to Friday, represented as Day 1 to Day 5 in later discussions) was selected from Autumn, Winter, and Spring for each of the classrooms (Fig. 11). Since the classrooms remained unoccupied for most of the summer vacation, this period was not included in our analysis. For each day, the data was selected from the first class start to the first $CO_2$ peak to do the evaluation. It is based on the assumption that the ventilation conditions remain the same for the whole day, and the number of students who attend the first class will be considered as the maximum attendance on that day. The ventilation rate and $CO_2$ emission rates were estimated using the developed approach, and occupancy was also calculated under the assumption that the average $CO_2$ generation rate per person was 0.0047 L/s [7].

Based on the estimated ventilation rates and occupancy levels, the equivalent clean airflow delivery rates per person were carried out (ECAi) and compared with the minimum values recommended by ASHRAE Standard 241 [34]. The ECAi sums the clean air supply rates contributed by indoor mitigation measures, including outdoor air ventilation, HVAC filtration, and air-cleaning devices such as portable air cleaners (PAC) or germicidal ultraviolet (GUV) [35-37]. This will assess the capability of the classroom to mitigate long-range aerosol exposures. In addition, it will help clarify the efforts required to achieve the infection risk management target established by ASHRAE 241 and figure out proper mitigation measures. The Equivalent Clean Air Calculator will be used for the assessment [34]. A steady-state $CO_2$ threshold that achieves minimum ECAi requirements was carried out for the evaluation periods and summarized with mean and pooled standard deviation for each classroom. Thresholds were established for scenarios involving pure ventilation, combined mitigation strategies, and various occupancy levels based on the summary of 2,000 runs of SDE model simulations for each scenario.

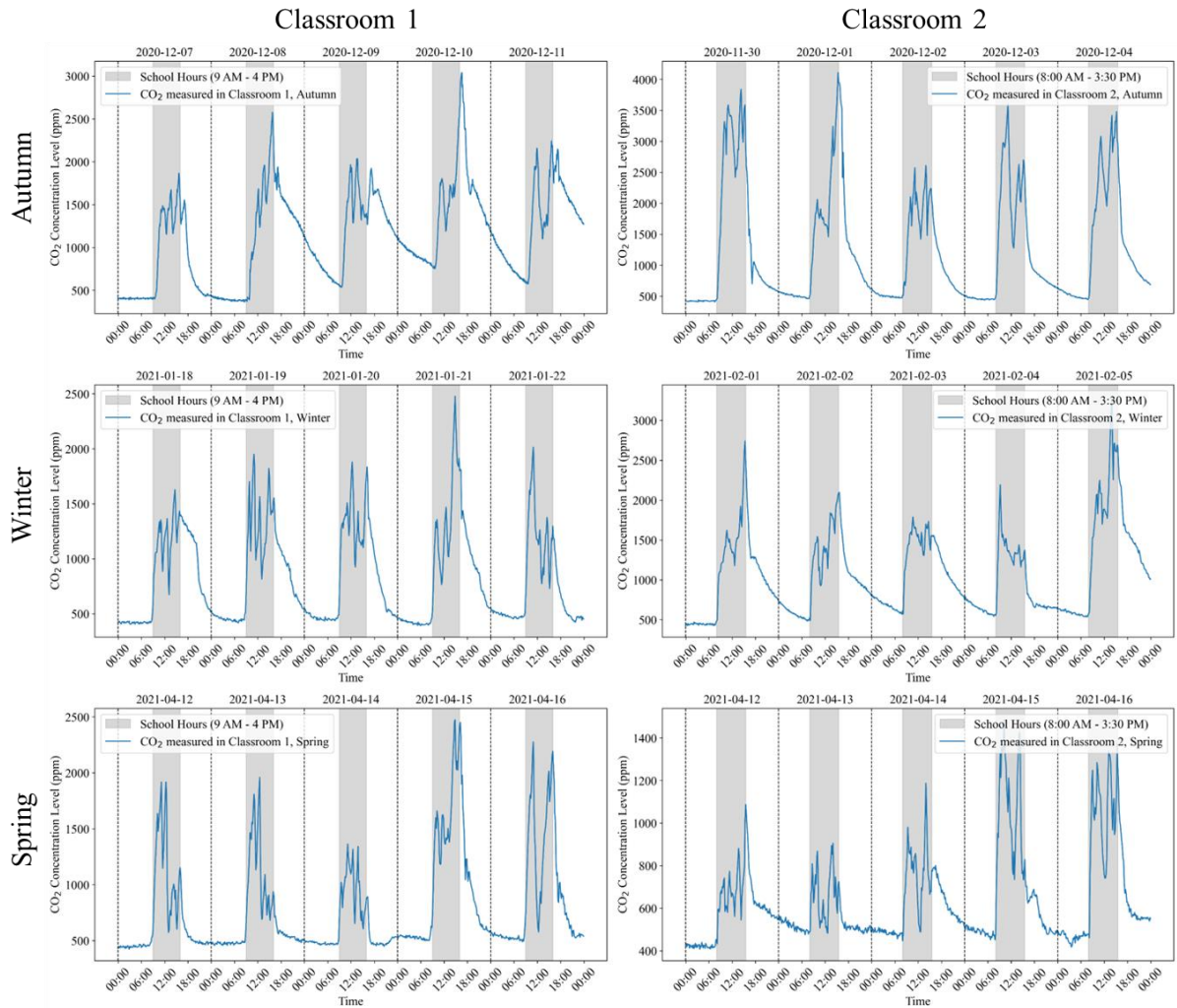

Fig. 11 One-week $CO_2$ measurements selected from Autumn, Winter, and Spring for two classrooms

### 3.3.1 Estimated parameters and ECAi

The proposed approach was subsequently employed to evaluate the ventilation conditions of two classrooms from two Montreal primary schools. As illustrated in Table 8, it could be found that the ventilation rates for Classroom 1 and Classroom 2 were inferred to be in the range of 0.11 – 1.38 ACH and 0.11 – 3.66 ACH, respectively. In evaluated days, the largest ventilation rate appears in Spring for both classrooms, which turns out to be Day 3, Spring (2021-04-14) for Classroom 1 and Day 2, Spring (2021-04-13) for Classroom 2. The occupancy turns out to be ranging from 9 - 20 for Classroom 1 and 14 - 20 for Classroom 2, with one exceptional day of only 3 students attending the class. The ECAi provided for each day was evaluated, ranging from 0.6 to 24.7 L/s/person. In the evaluated days, the ECAi provided in most of the days was significantly lower than the value recommended by ASHRAE 241 (20 L/s/person for Classroom) [34]. This suggests that, throughout the evaluation period, the clean air introduced into the two classrooms had limited capabilities in removing aerosols and failed to meet the ECAi requirements.

Table 8 Evaluation results for the two classrooms in Autumn, Winter, and Spring

| Classroom1, Autumn | Day1 | Day2 | Day3 | Day4 | Day5 |
|---|---|---|---|---|---|
| Ventilation rate (ACH) | 0.35 | 0.63 | 0.24 | 0.34 | 0.25 |
| Total $CO_2$ emission rate (L/s) | 0.044 | 0.079 | 0.09 | 0.083 | 0.088 |
| Estimated occupancy | 9 | 17 | 19 | 18 | 19 |
| ECAi provided (L/s/person) | 2.3 | 2.2 | 0.8 | 1.1 | 0.8 |
| Classroom1, Winter | Day1 | Day2 | Day3 | Day4 | Day5 |
| Ventilation rate (ACH) | 0.58 | 1.24 | 0.96 | 0.81 | 0.2 |
| Total $CO_2$ emission rate (L/s) | 0.044 | 0.069 | 0.078 | 0.081 | 0.091 |
| Estimated occupancy | 9 | 15 | 17 | 17 | 19 |
| ECAi provided (L/s person) | 3.9 | 5.0 | 3.4 | 2.9 | 0.6 |
| Classroom1, Spring | Day1 | Day2 | Day3 | Day4 | Day5 |
| Ventilation rate (ACH) | 0.11 | 0.44 | 1.38 | 0.48 | 0.14 |
| Total $CO_2$ emission rate (L/s) | 0.046 | 0.087 | 0.079 | 0.083 | 0.092 |
| Estimated occupancy | 10 | 19 | 17 | 18 | 20 |
| ECAi provided (L/s/person) | 0.7 | 1.4 | 4.9 | 1.6 | 0.4 |
| Classroom2, Autumn | Day1 | Day2 | Day3 | Day4 | Day5 |
| Ventilation rate (ACH) | 0.21 | 0.26 | 0.3 | 0.24 | 0.11 |
| Total $CO_2$ emission rate (L/s) | 0.084 | 0.089 | 0.085 | 0.083 | 0.092 |
| Estimated occupancy | 18 | 19 | 18 | 18 | 20 |
| ECAi provided (L/s/person) | 0.6 | 3.2 | 0.9 | 0.7 | 0.3 |
| Classroom2, Winter | Day1 | Day2 | Day3 | Day4 | Day5 |
| Ventilation rate (ACH) | 1.18 | 0.6 | 0.51 | 0.79 | 0.2 |
| Total $CO_2$ emission rate (L/s) | 0.071 | 0.083 | 0.085 | 0.071 | 0.09 |
| Estimated occupancy | 15 | 18 | 18 | 15 | 19 |
| ECAi provided (L/s/person) | 4.4 | 4.5 | 1.6 | 2.9 | 0.6 |
| Classroom2, Spring | Day1 | Day2 | Day3 | Day4 | Day5 |
| Ventilation rate (ACH) | 1.33 | 3.66 | 0.81 | 0.88 | 1.74 |
| Total $CO_2$ emission rate (L/s) | 0.015 | 0.068 | 0.086 | 0.081 | 0.069 |
| Estimated occupancy | 3 | 14 | 18 | 17 | 15 |
| ECAi provided (L/s/person) | 24.7 | 17.9 | 2.5 | 2.9 | 6.5 |

Note. The average $CO_2$ generation rate per person was assumed to be 0.0047 L/s for Classrooms (5-8 years) [7].

### 3.3.2 Steady-state $CO_2$ threshold achieving minimum ECAi requirements

To satisfy the minimum ECAi requirements, the steady-state $CO_2$ threshold was determined with the stochastic $CO_2$ grey-box model, employing outdoor ventilation as the only air-cleaning strategy. The minimum ECAi requirement of 20 L/s/person was used to determine the ventilation rate 'Q' in the model, alongside other parameters estimated from the previous evaluation phase. Two thousand $CO_2$ steady-state concentration simulations were conducted for each day evaluated in Table 8, and the summarized $CO_2$ steady-state concentration was shown in Fig. 12 for each of the classrooms (Classroom 1: 688.2 ± 132.4 ppm, Classroom 2: 690.3 ± 158.2 ppm). Daily evaluation results are shown in Appendix 1. The cumulative distribution of school-hour $CO_2$ measurements in Autumn, Winter, and Spring for Classroom 1 and Classroom 2 were also demonstrated in Fig. 12. In Classroom 1, only 25% of measurements fall into the established steady-state $CO_2$ threshold that achieves minimum ECAi requirements, while this number for Classroom 2 was 35%. This shows that during at least two-thirds of school hours, the minimum ECAi requirements are not met throughout the academic year. Natural ventilation alone is insufficient to ensure safe and healthy learning environments. Therefore, an increased supply of clean air in classrooms is necessary.

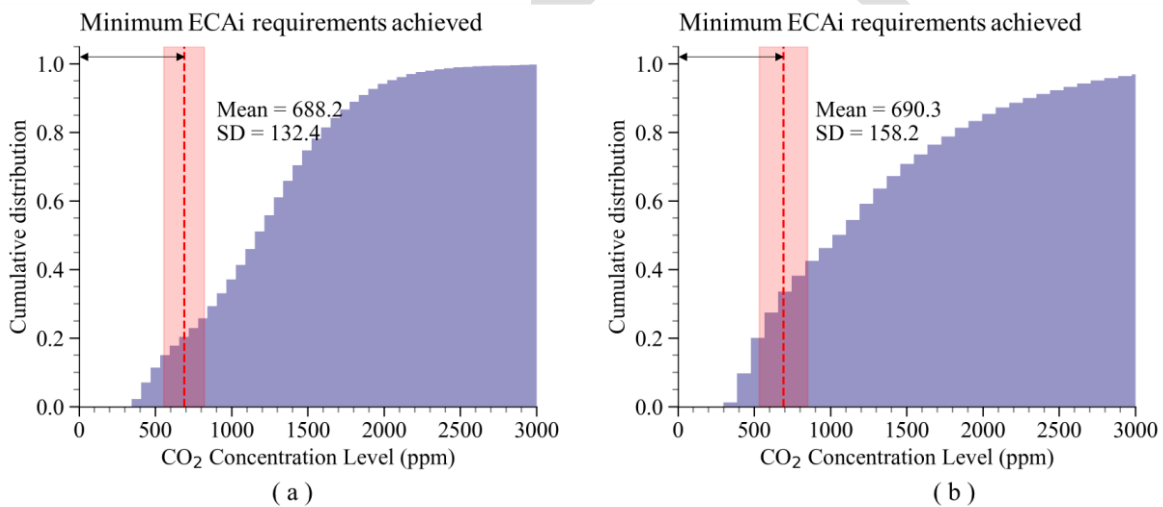

Fig. 12 Whole year $CO_2$ measurements in two classrooms; a) Cumulative distribution of school-hour $CO_2$ measurements in Autumn, Winter, and Spring for Classroom1; b) Cumulative distribution of school-hour $CO_2$ measurements in Autumn, Winter, and Spring for Classroom2

### 3.3.3 Retrofits to achieve the minimum ECAi required by ASHRAE 241

The impact of different exposure mitigation measures, with clean air delivery rates (CADR) ranging from 200 to 1000 cubic feet per minute (cfm), on ECAi was investigated, and the results are presented in Table 9. Various combinations of in-room UV devices and air cleaners (fan filter type) were evaluated to achieve varying levels of CADR. The findings suggest that a supplement of air-cleaning devices with a CADR of 800 cfm or more ensures that ECAi requirements are consistently achieved in both classrooms.

Table 9 ECAi under different mitigation measures

| Classroom1, Autumn | Day1 | Day2 | Day3 | Day4 | Day5 |
|---|---|---|---|---|---|
| In-room UV (200 CADR) | 12.8 | 7.8 | 5.7 | 6.4 | 5.8 |
| In-room air cleaner (400 CADR) | **23.3** | 13.3 | 10.7 | 11.6 | 10.7 |
| In-room UV + In-room air cleaner (600 CADR) | **33.8** | 18.9 | 15.7 | 16.9 | 15.7 |
| 2 × In-room air cleaner (800 CADR) | **44.3** | **24.4** | **20.6** | **22.1** | **20.7** |
| In-room UV + 2 × In-room air cleaner (1000 CADR) | **54.8** | **30.0** | **25.6** | **27.4** | **25.6** |
| Classroom1, Winter | Day1 | Day2 | Day3 | Day4 | Day5 |
| In-room UV (200 CADR) | 14.4 | 11.2 | 8.9 | 8.4 | 5.6 |
| In-room air cleaner (400 CADR) | **24.8** | 17.5 | 14.5 | 14.0 | 10.6 |
| In-room UV + In-room air cleaner (600 CADR) | **35.3** | **23.8** | **20.0** | 19.5 | 15.5 |
| 2 × In-room air cleaner (800 CADR) | **45.8** | **30.1** | **25.6** | **25.1** | **20.5** |
| In-room UV + 2 × In-room air cleaner (1000 CADR) | **56.3** | **36.4** | **31.2** | **30.6** | **25.5** |
| Classroom1, Spring | Day1 | Day2 | Day3 | Day4 | Day5 |
| In-room UV (200 CADR) | 10.1 | 6.4 | 10.4 | 6.8 | 5.1 |
| In-room air cleaner (400 CADR) | 19.5 | 11.3 | 16.0 | 12.1 | 9.9 |
| In-room UV + In-room air cleaner (600 CADR) | **29.0** | 16.3 | **21.5** | 17.3 | 14.6 |
| 2 × In-room air cleaner (800 CADR) | **38.4** | **21.3** | **27.1** | **22.6** | **19.3** |
| In-room UV + 2 × In-room air cleaner (1000 CADR) | **47.9** | **26.2** | **32.6** | **27.8** | **24.0** |
| Classroom2, Autumn | Day1 | Day2 | Day3 | Day4 | Day5 |
| In-room UV (200 CADR) | 5.9 | 8.2 | 6.2 | 6.0 | 5.0 |
| In-room air cleaner (400 CADR) | 11.1 | 13.2 | 11.4 | 11.2 | 9.7 |
| In-room UV + In-room air cleaner (600 CADR) | 16.4 | 18.2 | 16.7 | 16.5 | 14.5 |
| 2 × In-room air cleaner (800 CADR) | **21.6** | **23.1** | **21.9** | **21.7** | **19.2** |
| In-room UV + 2 × In-room air cleaner (1000 CADR) | **26.9** | **28.1** | **27.2** | **27.0** | **23.9** |
| Classroom2, Winter | Day1 | Day2 | Day3 | Day4 | Day5 |
| In-room UV (200 CADR) | 10.7 | 9.7 | 6.8 | 9.2 | 5.6 |
| In-room air cleaner (400 CADR) | 17.0 | 15.0 | 12.1 | 15.5 | 10.5 |
| In-room UV + In-room air cleaner (600 CADR) | **23.3** | **20.2** | 17.3 | **21.8** | 15.5 |
| 2 × In-room air cleaner (800 CADR) | **29.6** | **25.5** | **22.6** | **28.1** | **20.5** |
| In-room UV + 2 × In-room air cleaner (1000 CADR) | **35.8** | **30.7** | **27.8** | **34.4** | **25.4** |
| Classroom2, Spring | Day1 | Day2 | Day3 | Day4 | Day5 |
| In-room UV (200 CADR) | **56.2** | **24.7** | 7.7 | 8.4 | 12.8 |
| In-room air cleaner (400 CADR) | **87.6** | **31.4** | 13.0 | 14.0 | 19.0 |
| In-room UV + In-room air cleaner (600 CADR) | **119.1** | **38.2** | 18.2 | 19.5 | **25.3** |
| 2 × In-room air cleaner (800 CADR) | **150.6** | **44.9** | **23.5** | **25.1** | **31.6** |
| In-room UV + 2 × In-room air cleaner (1000 CADR) | **182.0** | **51.6** | **28.7** | **30.6** | **37.9** |

Note: The conditions that satisfy the EACi requirements in ASHRAE 241 (20 L/s/person) are in bold

### 3.3.4 Manage long-range indoor aerosol exposures using $CO_2$ as a proxy

For the purpose of creating a clean and healthy indoor environment, the $CO_2$ thresholds were established for indoor ventilation designs and operations in the classrooms. The steady-state $CO_2$ levels were carried out using the stochastic $CO_2$-based grey-box model. Three aerosol exposure management levels were established from two thousand predictive outcomes of the model: $C_{limit}$ (Mean + SD) as the maximum threshold indicating poor ventilation beyond this limit, $C_{target}$ (Mean) as the expected $CO_2$ concentration limit, under which conditions are deemed acceptable and generally comply with ECAi, and $C_{ideal}$ (Mean – SD) as the optimal threshold, recommended when infection risk of respiratory diseases in the classroom is a significant concern. These thresholds could help manage long-range indoor aerosol exposures by using $CO_2$ as a proxy while taking real-life uncertainties into consideration.

The design $C_{target}$ levels were evaluated for classrooms under varying occupancy and CADR conditions (Fig. 13). When no additional CADR is supplied, a ventilation rate of 20 L/s per person is required, resulting in an average $C_{target}$ level of 683 ppm and 686 ppm for the classrooms respectively. Thus, it is suggested to set $C_{target}$ below 690 ppm when managing indoor aerosol exposures is a priority. In scenarios where air-cleaning devices with sufficient CADR are adopted (800 cfm), the $C_{target}$ stabilizes around 1000 ppm. Conversely, when limited CADR is supplemented such as 200 cfm, the $C_{target}$ level initially rises with increased occupancy but subsequently falls as additional ventilation is needed to maintain the effective clean air level (ECAi).

Uncertainties in real-life operations can influence the estimated maximum $CO_2$ levels used to indicate whether a room meets ECAi requirements. For instance, actual attendance may vary from the designed occupancy levels. As a result, the $CO_2$ thresholds ($C_{limit}$, $C_{target}$, and $C_{ideal}$) were carried out for different mitigation measures with CADR ranging from 200 to 1000 cfm. These thresholds are depicted in Fig. 14 for Classroom 1 and Classroom 2, and generalized equations derived from the average of their coefficients are presented in Eq. 6 to Eq. 8. When the CADR is below 600 cfm, enhancing air-cleaning capacity improves ECAi, thereby reducing reliance on outdoor ventilation for achieving ECAi requirements. While the contribution from outdoor ventilation can decrease from the initial 20 L/s/person, it must still meet the minimum ventilation rate of 7.4 L/s/person recommended in ASHRAE 62.1 for classrooms [13]. Once air-cleaning devices provide sufficient ECAi, the steady-state $CO_2$ thresholds indicating ECAi satisfaction remain stable.

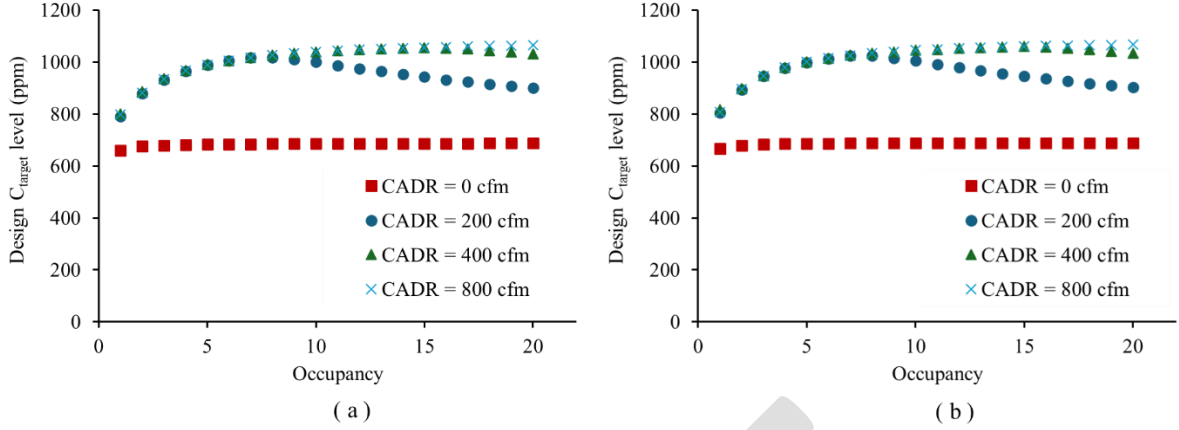

Fig. 13 Design $C_{target}$ level for different occupancy in two classrooms; a) Classroom 1; b) Classroom 2

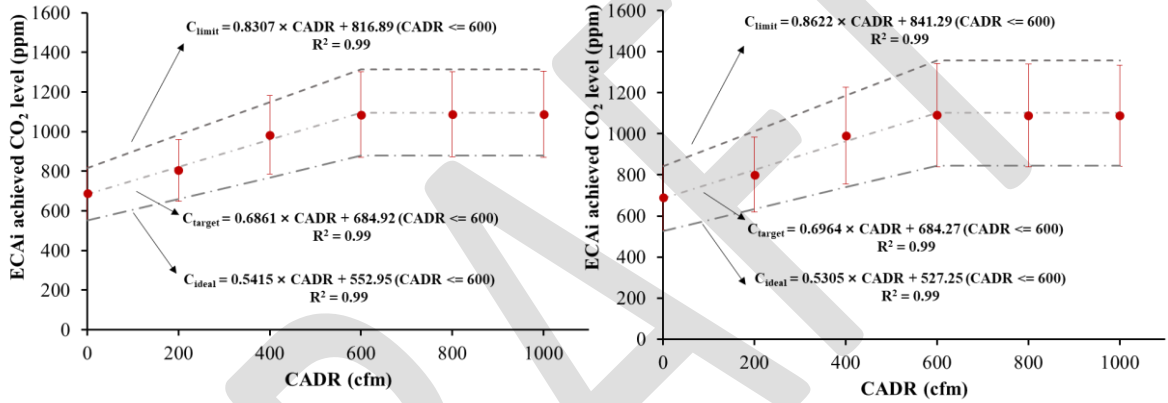

Fig. 14 Steady-state $CO_2$ thresholds that achieve minimum ECAi requirements with the employment of air-cleaning devices under different CADR; a) Classroom 1; b) Classroom 2

$$C_{limit} = \begin{cases} 0.8 \times CADR + 829.1 & (CADR \leq 600) \\ 1309.1 & (CADR > 600) \end{cases} \quad 6)$$

$$C_{target} = \begin{cases} 0.7 \times CADR + 684.6 & (CADR \leq 600) \\ 1104.6 & (CADR > 600) \end{cases} \quad 7)$$

$$C_{ideal} = \begin{cases} 0.5 \times CADR + 540.1 & (CADR \leq 600) \\ 840.1 & (CADR > 600) \end{cases} \quad 8)$$

When the CADR level of air-cleaning devices introduced into classrooms is set, the $C_{limit}$, $C_{target}$, and $C_{ideal}$ can be respectively calculated for classrooms with similar designs as the two Montreal primary classrooms investigated in this study. The 'similar designs' refer to aspects such as dimensions, occupancy, attendance, ventilation mode, etc. For other public indoor facilities with distinctive ventilation designs, measuring $CO_2$ concentrations during occupied hours is advised to obtain the data for inference. Subsequently, case-specific $CO_2$ thresholds can be determined using the methodology established in this study.

It should also be noted that the $CO_2$ thresholds established here aim at indicating whether the IAQ in classrooms complies with ASHRAE standard 241 [34] and ASHRAE standard 62.1 [13]. ASHRAE standard 241 outlines the clean-air requirements within an Infection Risk Management Mode (IRMM) during an outbreak, requiring higher cleaning air delivery levels when compared with ASHRAE standard 62.1. In scenarios where air-cleaning devices are absent or limited, a considerable volume of outdoor ventilation is recommended to maintain air quality (e.g., when the CADR is 0, a ventilation rate of 20 L/s/person is advised). As the CADR available to the room increases, the requirement for outdoor ventilation decreases accordingly. Nonetheless, it needs to be realized that even when air-cleaning devices supply sufficient ECAi, the outdoor ventilation rates must still adhere to the minimum requirements outlined in ASHRAE standard 62.1 to ensure adequate air quality (for instance, when CADR is equal to or greater than 600 cfm, a minimum ventilation rate of 7.4 L/s/person is still mandated). When the community infection risk of respiratory diseases is low, it is also appropriate to utilize the thresholds designed for scenarios with sufficient CADR, such as during the plateau periods when the CADR exceeds 600. These thresholds align with the ventilation requirements specified by ASHRAE Standard 62.1 only.

## 4. Conclusions

In this study, we provide an innovative approach to quantify uncertainties in indoor ventilation condition evaluations of Canadian primary schools. The approach proposed by this study can help interpret $CO_2$ recordings in real classroom settings and predict steady-state $CO_2$ levels considering uncertainties. Here are the main contributions of this study:

- By employing Bayesian inference on a $CO_2$-based grey-box SDE model, the ventilation rate and $CO_2$ emission rate can be accurately predicted. Uncertainties come from measurements, the randomness of air movements, and modeled or unmodelled parameters, which can be quantified using the incremental variance $\sigma$.
- The robustness and reliability of the model were validated with $CO_2$ tracer gas experiments in an airtight chamber. Prior sensitivity analysis was conducted to verify the rationality of assumed prior assumptions. Parameters inferred from the model were compared with chamber measurements to confirm its estimation accuracy. The PPC evaluations were conducted to see whether the estimated parameters for the model could work well to represent the observations. The results suggested that the model is robust to its prior assumptions and can estimate the interested parameters with reliable accuracy.

- Applications were conducted to interpret the real-life $CO_2$ measurements in two classrooms in Montreal. Using the estimated ventilation and occupancy, the provided ECAi and the steady-state $CO_2$ threshold for achieving minimum ECAi requirements were calculated, suggesting natural ventilation is insufficient to achieve ECAi standards established by ASHRAE 241 for all three seasons.
- Adopting a CADR of 800 cfm can help the classrooms to effectively manage aerosol exposures. In addition, steady-state $CO_2$ thresholds ($C_{limit}$, $C_{target}$, and $C_{ideal}$) to indicate the ECAi satisfactory status were carried out for different mitigations using the stochastic $CO_2$ grey-box model and inferred parameters.
- To achieve the minimum ECAi level required by ASHRAE 241, the target $CO_2$ level is suggested to be below 690 ppm for similar classrooms without additional clean-air treatment. When sufficient clean air is supplemented, the design $C_{target}$ is appropriately set at 1000 ppm. Empirical equations were also established for classrooms that share the ventilation design featured in this study. In real-life operations and management, it is recommended to reference Fig. 14 to consider uncertainties.

Limitations do exist in this study for only two classrooms were investigated, and occupancy information was not available for further verifications of the model. For people who are interested in understanding the ECAi-compliant steady-state $CO_2$ thresholds for a specific indoor environment, it is advisable to conduct consistent field $CO_2$ measurements in occupied hours for the intended scenario and apply the approach developed in this study.

**Acknowledgments**

This work was supported by the National Research Council Canada contract [#980615 and #999657] through "Benchmarking test facilities for capturing aerosol movement in building spaces – an evidence-centered study (Phase I and II)". The long-term $CO_2$ measurement data were provided from the Natural Sciences and Engineering Research Council of Canada (NSERC) project "Assessment and mitigation of summertime overheating conditions in vulnerable buildings of urban agglomerations" [#ACCPJ 535986-18] under the Advancing Climate Change Science in Canada Program.

Appendix1 The daily steady-state $CO_2$ threshold for achieving minimum ECAi requirements

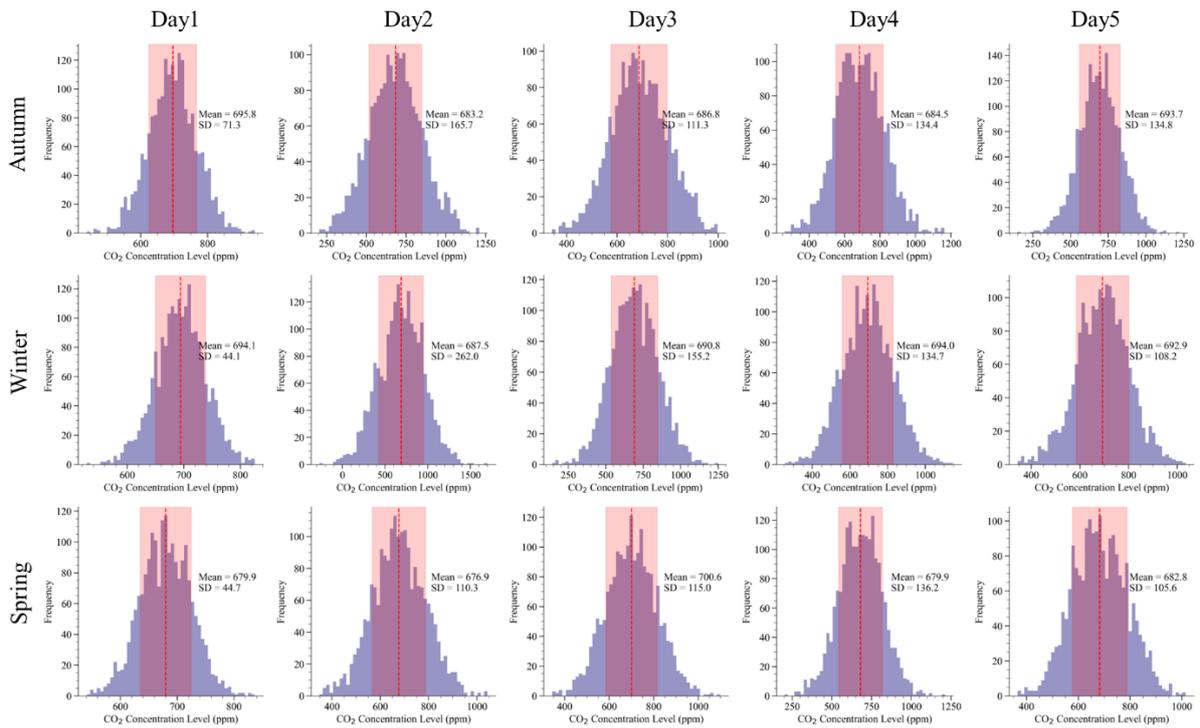

Fig.A 1 The selected daily steady-state $CO_2$ threshold evaluated for Classroom 1 to achieve minimum ECAi requirements

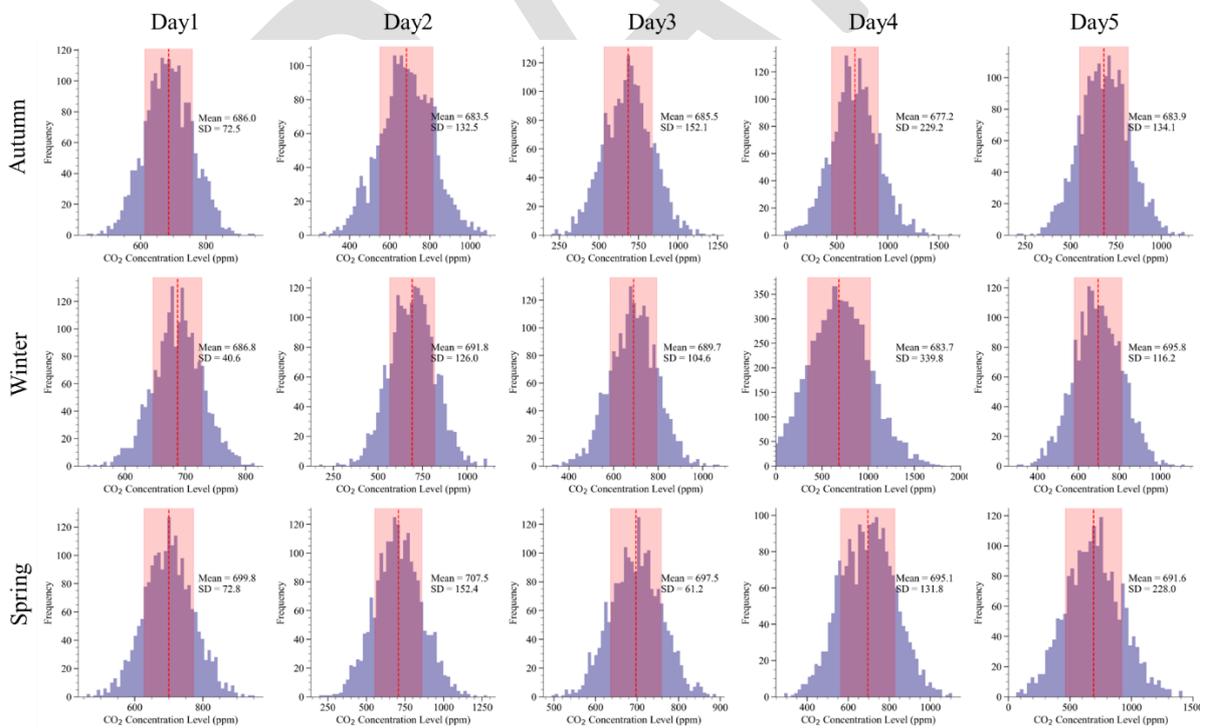

Fig.A 2 The selected daily steady-state $CO_2$ threshold evaluated for Classroom 2 to achieve minimum ECAi requirements

**References**

[1] "WHO Coronavirus (COVID-19) dashboard " World Health Organization 2023 data. https://data.who.int/dashboards/covid19/data (accessed April 7, 2024).

[2] C. C. Wang et al., "Airborne transmission of respiratory viruses," *Science,* vol. 373, no. 6558, p. eabd9149, doi: 10.1126/science.abd9149.

[3] L. Morawska et al., "How can airborne transmission of COVID-19 indoors be minimised?," *Environment international,* vol. 142, p. 105832, 2020.

[4] I. Garg, R. Shekhar, A. B. Sheikh, and S. Pal, "Impact of COVID-19 on the changing patterns of respiratory syncytial virus infections," *Infectious Disease Reports,* vol. 14, no. 4, pp. 558-568, 2022.

[5] L. Chatzidiakou, D. Mumovic, and A. J. Summerfield, "What do we know about indoor air quality in school classrooms? A critical review of the literature," *Intelligent Buildings International,* vol. 4, no. 4, pp. 228-259, 2012.

[6] *ASHRAE Position Document on Indoor Carbon Dioxide*, ASHRAE, 2022.

[7] A. Persily, "Development and application of an indoor carbon dioxide metric," *Indoor Air,* vol. 32, no. 7, p. e13059, 2022.

[8] M. St-Jean et al., "Indoor air quality in Montréal area day-care centres, Canada," *Environmental Research,* vol. 118, pp. 1-7, 2012/10/01/ 2012, doi: https://doi.org/10.1016/j.envres.2012.07.001.

[9] M. M. Andamon, P. Rajagopalan, and J. Woo, "Evaluation of ventilation in Australian school classrooms using long-term indoor $CO_2$ concentration measurements," *Building and Environment,* vol. 237, p. 110313, 2023/06/01/ 2023, doi: https://doi.org/10.1016/j.buildenv.2023.110313.

[10] J. Lofaro. "Education ministry says 90,000 $CO_2$ readers to be installed in classrooms across Quebec." CTVNewsMontreal. https://montreal.ctvnews.ca/education-ministry-says-90-000-co2-readers-to-be-installed-in-classrooms-across-quebec-1.5581304 (accessed 04/16, 2024).

[11] A. Kabirikopaei and J. Lau, "Uncertainty analysis of various CO2-Based tracer-gas methods for estimating seasonal ventilation rates in classrooms with different mechanical systems," *Building and Environment,* vol. 179, p. 107003, 2020/07/15/ 2020, doi: https://doi.org/10.1016/j.buildenv.2020.107003.

[12] S. Batterman, "Review and extension of CO2-based methods to determine ventilation rates with application to school classrooms," *International journal of environmental research and public health,* vol. 14, no. 2, p. 145, 2017.

[13] *Ventilation and Acceptable Indoor Air Quality*, ASHRAE-62.1, 2022.

[14] T. P. Bohlin, *Practical grey-box process identification: theory and applications*. Springer Science & Business Media, 2006.

[15] M. Macarulla, M. Casals, M. Carnevali, N. Forcada, and M. Gangolells, "Modelling indoor air carbon dioxide concentration using grey-box models," *Building and Environment,* vol. 117, pp. 146-153, 2017.

[16] M. Macarulla, M. Casals, N. Forcada, M. Gangolells, and A. Giretti, "Estimation of a room ventilation air change rate using a stochastic grey-box modelling approach," *Measurement,* vol. 124, pp. 539-548, 2018.

[17] F. Haghighat, P. Fazio, and T. Unny, "A predictive stochastic model for indoor air quality," *Building and Environment,* vol. 23, no. 3, pp. 195-201, 1988.

[18] B. K. Øksendal, *Stochastic differential equations : an introduction with applications*, 6th , corr. ed. (Universitext). Berlin ;: Springer, 2007.

[19] F. Wang, X. Zhou, and H. Kikumoto, "Improvement of optimization methods in indoor time-variant source parameters estimation combining unsteady adjoint equations and flow field information," *Building and Environment,* vol. 226, p. 109710, 2022.

[20] F. Wang, X. Zhou, J. Huang, H. Wang, H. Kikumoto, and C. Deng, "Natural gas leakage estimation in underground utility tunnels using Bayesian inference based on flow fields with gas jet disturbance," *Process Safety and Environmental Protection,* vol. 165, pp. 532-544, 2022.

[21] F. Septier, P. Armand, and C. Duchenne, "A bayesian inference procedure based on inverse dispersion modelling for source term estimation in built-up environments," *Atmospheric Environment,* vol. 242, p. 117733, 2020.

[22] D. Hou *et al.*, "Development of a Bayesian inference model for assessing ventilation condition based on CO2 meters in primary schools," in *Building simulation*, 2023, vol. 16, no. 1: Springer, pp. 133-149.

[23] H. Rahman and H. Han, "Bayesian estimation of occupancy distribution in a multi-room office building based on $CO_2$ concentrations," in *Building Simulation*, 2018, vol. 11: Springer, pp. 575-583.

[24] S. Madhira and S. Deshmukh, "Brownian Motion Process," in *Introduction to Stochastic Processes Using R*: Springer, 2023, pp. 487-545.

[25] X. Mao, "The truncated Euler–Maruyama method for stochastic differential equations," *Journal of Computational and Applied Mathematics,* vol. 290, pp. 370-384, 2015.

[26] Y. Zhong, A. D. Knefaty, G. Chen, J. Yao, and R. Zheng, "Forecast of air-conditioning duration in office buildings in summer using machine learning and Bayesian theories," *Journal of Building Engineering,* vol. 61, p. 105218, 2022.

[27] T. Zhao, J. Li, P. Wang, S. Yoon, and J. Wang, "Improvement of virtual in-situ calibration in air handling unit using data preprocessing based on Gaussian mixture model," *Energy and Buildings,* vol. 256, p. 111735, 2022.

[28] Z. Wang and T. Hong, "Learning occupants' indoor comfort temperature through a Bayesian inference approach for office buildings in United States," *Renewable and Sustainable Energy Reviews,* vol. 119, p. 109593, 2020.

[29] M. Richey, "The evolution of Markov chain Monte Carlo methods," *The American Mathematical Monthly,* vol. 117, no. 5, pp. 383-413, 2010.

[30] A. Gelman, J. B. Carlin, H. S. Stern, and D. B. Rubin, *Bayesian data analysis*. Chapman and Hall/CRC, 1995.

[31] J. Rockström *et al.*, "Planetary Boundaries Exploring the Safe Operating Space for Humanity," *Ecology and Society,* vol. 14, no. 2, 2009. [Online]. Available: http://www.jstor.org/stable/26268316.

[32] A. Patil, D. Huard, and C. J. Fonnesbeck, "PyMC: Bayesian stochastic modelling in Python," *Journal of statistical software,* vol. 35, no. 4, p. 1, 2010.

[33] *Thermal environmental conditions for human occupancy*, ASHRAE-55, 2023.

[34] *Control of Infectious Aerosols*, ASHRAE-241, 2023.

[35] S. Yan, L. Wang, M. J. Birnkrant, J. Zhai, and S. L. Miller, "Evaluating SARS-CoV-2 airborne quanta transmission and exposure risk in a mechanically ventilated multizone office building," *Building and Environment,* vol. 219, p. 109184, 2022/07/01/ 2022, doi: https://doi.org/10.1016/j.buildenv.2022.109184.

[36] S. Yan, L. Wang, M. J. Birnkrant, Z. Zhai, and S. L. Miller, "Multizone Modeling of Airborne SARS-CoV-2 Quanta Transmission and Infection Mitigation Strategies in Office, Hotel, Retail, and School Buildings," *Buildings*, vol. 13, no. 1*,* doi: 10.3390/buildings13010102.

[37] S. Rayegan *et al.*, "A review on indoor airborne transmission of COVID-19– modelling and mitigation approaches," *Journal of Building Engineering,* vol. 64, p. 105599, 2023/04/01/ 2023, doi: https://doi.org/10.1016/j.jobe.2022.105599.